\documentclass{aa}

\usepackage[varg]{txfonts}
\usepackage{graphicx}
\usepackage[usenames,dvipsnames,svgnames,table]{xcolor}
\usepackage[colorlinks=true,citecolor=blue]{hyperref}
\usepackage{soul}
\usepackage{enumitem}

\begin{document}

\title{Cluster counts. II. Tensions, massive neutrinos, and modified gravity}
\titlerunning{Cluster counts. II.}
\author{Stéphane Ili\'c\inst{1,2}
\and Ziad Sakr\inst{2,3}
\and Alain Blanchard\inst{2}}
\authorrunning{Ili\'c et al.}
\institute{
    CEICO, Institute of Physics of the Czech Academy of Sciences, Na Slovance 2, Praha 8 Czech Republic
    \and
    IRAP, Université de Toulouse, CNRS, CNES, UPS, Toulouse, France
    \and
    Universit\'e St Joseph; UR EGFEM, Faculty of Sciences, Beirut, Lebanon
}
\date{Received XXXX-XX-XX; accepted XXXX-XX-XX}

\abstract{
  The Lambda cold dark matter ($\Lambda$CDM) concordance model is very successful at describing our Universe with high accuracy and only a few parameters. Despite its successes, a few tensions persist; most notably, the best-fit $\Lambda$CDM model, as derived from the Planck cosmic microwave background (CMB) data, largely overpredicts the abundance of Sunyaev-Zel'dovich (SZ) clusters when using their standard mass calibration. Whether this is the sign of an incorrect calibration or the need for new physics remains a matter of debate. In this study, we examined two simple extensions of the standard model and their ability to release the aforementioned tension: massive neutrinos and a simple modified gravity model via a non-standard growth index $\gamma$. We used both the Planck CMB power spectra and SZ cluster counts as datasets, alone and in combination with local X-ray clusters. In the case of massive neutrinos, the cluster-mass calibration $(1-b)$ is constrained to $0.585^{+0.031}_{-0.037}$ (68\% limits), more than 5$\sigma$ away from its standard value $(1-b)\sim 0.8$. We found little correlation between neutrino masses and cluster calibration, corroborating previous conclusions derived from X-ray clusters; massive neutrinos do not alleviate the cluster-CMB tension. With our simple $\gamma$ model, we found a large correlation between the calibration and the growth index $\gamma,$ but contrary to local X-ray clusters, SZ clusters are able to break the degeneracy between the two parameters thanks to their extended redshift range. The calibration $(1-b)$  was then constrained to $0.602^{+0.053}_{-0.065}$, leading to an interesting constraint on $\gamma = 0.60\pm 0.13$. When both massive neutrinos and modified gravity were allowed, preferred values remained centred on standard $\Lambda$CDM values, but a calibration $(1-b)\sim 0.8$ was allowed (though only at the $2\sigma$ level) provided $\sum m_\nu \sim 0.34 $ eV and $\gamma \sim 0.8$. We conclude that massive neutrinos do not relieve the cluster-CMB tension, and that a calibration close to the standard value $(1-b)\sim 0.8$ would call for new physics in the gravitational sector.
}


\maketitle

\section{Introduction}\label{sec:1-Intro}

The Lambda cold dark matter ($\Lambda$CDM) model is remarkably successful at describing the majority of observations relevant to cosmology. These include probes of the background evolution of our Universe, its early times, and the dynamics and evolution of matter perturbations. Perhaps the most striking success of the $\Lambda$CDM model is its ability to explain the observed fluctuations of the cosmic microwave background (CMB), as measured most notably by the Planck satellite, which led to a determination of the cosmological parameters in the $\Lambda$CDM model with unprecedented accuracy.

However, some tension is also noted, particularly between the $\Lambda$CDM cosmological parameters favoured by Sun\-ya\-ev\--\-Zel'\-do\-vich (SZ) cluster counts, and those derived from the angular power spectra of CMB temperature and polarisation fluctuations \citep{2014A&A...571A..20P,2016A&A...594A..24P}. This particular tension appears only when the standard mass calibration $(1-b)=0.8$ of SZ clusters (defined as the ratio of the hydrostatic to true mass) is used, leading to a significant discrepancy in the $\sigma_8$ (the present-day amplitude of matter fluctuations) versus $\Omega_m$ (the dark matter density) plane, or equivalently a discrepancy on the calibration parameter when left free, compared to its standard value \citep[see Fig.~3 in][and Fig.~\ref{fig:om_vs_s8} in the present article]{2018A&A...620A..78S}.

A critical aspect of this tension is that the mass of a cluster (including its dark matter halo) is not directly observable. The above tension is therefore present only when additional priors on cluster masses are used, and it more generally relies on the use of scaling relations between halo mass, redshift, and the observable cluster properties (known as mass proxies). Based on the assumption of hydrostatic equilibrium, the Planck Collaboration used X-ray observations from XMM-Newton in their analysis to derive masses of the intra-cluster medium \citep{2010A&A...517A..92A} and corrected the value by 20\% to account for the incomplete thermalisation of the gas. This correction is expressed as a `bias' $b$ in mass, defined as $(1-b)=0.8$ in the fiducial case.

Should the aforementioned tension be confirmed, we would be forced to consider extensions or alternatives to the standard $\Lambda$CDM model of cosmology. Since the CMB itself is mostly sensitive to the physics of the early Universe, one can reconcile it with cluster observations by introducing a modification to the cosmological model that has a significant impact preferentially at late times. More specifically, a new theory leading to a lower growth rate of structures would be required in order to predict a lower abundance of clusters, which would better reflect the data.

Such growth can be achieved in several ways; one possibility is to add mass to neutrinos in the standard cosmology (beyond their minimal experimental mass, 0.06 eV). Among other effects, massive neutrinos slow down the growth of matter perturbations during the matter and dark energy-dominated eras on scales smaller than their free-streaming length \citep[see][for a review on the effect of neutrinos in cosmology]{2012arXiv1212.6154L}. Alternatively, one could consider replacing the standard cosmological constant with a different form of dark energy. A fluid with a varying equation of state (the so-called quintessence models) or the addition of a scalar field can achieve the desired result, but a more radical approach consists of modifying the current theory of general relativity itself \citep[see][for extensive reviews]{2012PhR...513....1C,2019LRR....22....1I}.

In \citet{2018A&A...620A..78S}, we pointed out a similar discrepancy between the abundance of local X-ray clusters and CMB cosmology. We found that massive neutrinos do not solve the issue, while a simple modified gravity model parametrised by a modified growth rate, the $\gamma$ model, can reconcile the Planck cluster mass calibration with the Planck CMB cosmology at the expense of a relatively high $\gamma \sim 0.9$ \citep[a value potentially already excluded by current data; see ][]{2018MNRAS.476.3263L,2018MNRAS.477.1639Z}. In our study, we combined constraints from the redshift-mass distribution of SZ clusters with the latest CMB measurements from the Planck satellite, performing a Bayesian analysis through a Markov chain Monte Carlo (MCMC) analysis. We varied the parameters of the standard model, the cluster mass calibration $(1-b)$, and the additional parameters introduced in the context of the extended models we considered. They included the total mass of three massive degenerate neutrinos and the phenomenological $\gamma$ parameter of our simple modified gravity model.

In Sect.~\ref{sec:2-Meth}, we describe the formalism used for predicting cluster abundances, as well as the extensions to the standard model we examined. In the same section, we also detail the datasets used in this work, and the implementation of the MCMC analysis which samples the posterior probability distribution function. We present and discuss our results in Sect.~\ref{sec:3-Res}, while we summarise and discuss our conclusions in Sect.~\ref{sec:4-CCL}.

\section{Methodology and data}\label{sec:2-Meth}

\subsection{Cluster abundances as a cosmological probe}\label{ssec:2.1-Clus}

Clusters of galaxies provide tight constraints on cosmological models, as they (mostly) probe the growth of matter fluctuations directly throughout the history of the Universe. Their use as cosmological probes has been extensively studied and exploited in the past. The methodology that we adopted here is closely related to the approach followed by \citet{2015A&A...582A..79I} and \citet{2018A&A...620A..78S}. We refer the interested reader to those two articles for more details, as we only provide the main steps here.

We computed the expected numbers of clusters (as a function of mass and redshift) in a given cosmological model via integrals over the mass function, with the latter being taken from \citet{2016MNRAS.456.2486D} (thereafter D16). This more recent mass function was chosen over available alternatives \citep[notably the popular][mass function]{2008ApJ...688..709T}, although this choice has only a minimal impact on our results and conclusions, as can be seen in Fig.~\ref{fig:om_vs_s8} when comparing the filled (D16) and dash-dotted (Tinker) purple contours. The D16 mass function was optimised to work with the virial radius as a definition for clusters, which is the convention we used for X-ray clusters, although this choice had little impact on observables when modelling was done self-consistently \citep[see][]{2018A&A...620A..78S}. We did, however, use the `critical $M_{500}$' convention (i.e. the density contrast of clusters is defined as 500 times the critical density of the Universe) for SZ clusters, as the available Planck SZ dataset adopts this particular convention. We used the analytical formulae provided by the D16 authors in order to convert their mass function between their fiducial (virial) cluster mass definition and any other definition (critical $M_{500}$ here).

In the case of massive neutrinos, we used the so-called CDM prescription to amend the mass function \citep{2013JCAP...12..012C,2014JCAP...02..049C} unless otherwise stated. The filled and dashed orange contours of Fig.~\ref{fig:om_vs_s8} illustrate the effect of the CDM prescription in the analysis; as discussed in \citet{2018A&A...620A..78S}, we include its effects for the sake of completeness, even though it led to only marginal differences. As seen in Fig.~\ref{fig:om_vs_s8}, it led to slightly lower values of $\sigma_8,$ and concurrently high values of $\Omega_m,$ but with slightly increased error bars on both.

When considering a simple modified gravity model, as originally proposed by \citet{2005PhRvD..72d3529L}, we modelled the linear growth rate of structures like so\begin{equation}\label{eq:omgam}
f(a)=\Omega_m(a)^\gamma, 
\end{equation}
where $\gamma$, the growth index, is here a free parameter generalising the approximation $\gamma \sim 0.545$ in standard gravity \citep{1980lssu.book.....P}. The growth rate $f$ is defined as $f(a)\equiv d\ln G(a)/d\ln a$, where $G$ is the (scale-independent) linear growth factor of matter density perturbations $\delta_m$ defined as $\delta_m(a) = \delta_{m,0} G(a)/G(0)$.

In all extensions to the standard model considered here, we assume the non-linear regime would follow the same threshold $\delta_c,$ and the same virial density, $\Delta_v,$ as in the standard picture. Scaling laws are then used to relate theoretical cluster quantities (i.e. mass) to observable quantities. For the X-ray temperature $T$ versus cluster mass $M$, we use the standard scaling relation
\begin{equation}\label{eq:scallaw-X}
  T=A_{T-M}(h\, M_\Delta)^{2/3}\left(\frac{\Omega_{m} \Delta(z)}{178}\right)^{1/3}(1+z)
,\end{equation}
where $A_{T-M}$ is the normalisation parameter, $\Delta$ is the density contrast chosen for the definition of a cluster, expressed with respect to the total background matter density\footnote{Eq.~(\ref{eq:scallaw-X}) is different when the contrast density $\Delta_c$ is expressed with respect to the critical density at redshift $z$.} of the Universe at redshift $z$, and $M_\Delta$ is the mass of the cluster according to the same definition (as mentioned before, in our case for X-ray clusters, we used the virial radius and mass definition). Similarly, we used the following scaling relation between the expected mean SZ signal (denoted by the letter $\bar{Y}$) and mass $M_{500}$
\begin{equation}\label{eq:scallaw-SZ} 
E^{-\beta}(z)\left[\frac{D_A^2(z) \bar{Y}_{500}}{\mathrm{10^{-4}\,Mpc^2}}\right] =  Y_\ast \left[ \frac{h}{0.7}
  \right]^{-2+\alpha} \left[\frac{(1-b)\,
    M_{500}}{6\times10^{14}\,M_\odot}\right]^{\alpha},
\end{equation}
where the quantity $D_A$ refers to the angular diameter distance and $E(z)\equiv H(z)/H_0$. We note here the aforementioned hydrostatic mass bias $(1-b)$, while $\alpha$, $\beta$, and $Y_\ast$ are additional free parameters in the SZ scaling law. A fourth free parameter exists, namely $\sigma_{\ln Y}$; it corresponds to the scatter of the log-normal distribution that the measured $Y_{500}$ is assumed to follow (and of which the mean $\bar{Y}_{500}$ is given by Eq.~(\ref{eq:scallaw-SZ}))\footnote{\citet{2016A&A...594A..24P} provides best-fit values and uncertainties for all four of those parameters, derived from a fit based on X-ray observations. In our analysis, we follow the conventions used by the Planck Collaboration by fixing the $\beta$ parameter to its best-fit value ($=2/3$), letting $\sigma_{\ln Y}$ free with Gaussian priors derived from its fit ($=0.075\pm0.01$), and leaving $\alpha$ completely free. We do, however, fix $Y_\ast$ to its best-fit value ($\log Y_\ast = -0.186$) instead of leaving it free, noting that it is degenerate with the bias parameter $(1-b)$.}.

\subsection{Datasets and numerical tools}\label{ssec:2.2-Data}

We followed a standard MCMC analysis using two publicly available codes: the CosmoMC package \citep{2002PhRvD..66j3511L, 2013PhRvD..87j3529L} and the Monte Python code \citep{Audren:2012wb} that respectively use the CAMB and CLASS Boltzmann codes for the computation of relevant cosmological quantities: the matter power spectrum and the power spectra of CMB anisotropies. Those two codes were used to explore our full-parameter space under the constraint of our data sets extracted from CMB measurements \citep{2016A&A...594A..11P}, baryon acoustic oscillations (BAO) data \citep{2014MNRAS.441...24A,2014JCAP...05..027F}, X-ray-selected clusters \citep{2015A&A...582A..79I} and an SZ clusters sample \citep{2016A&A...594A..24P}, separated or  combined.

The 2015 Planck CMB \citep{2016A&A...594A...1P} and BAO \citep{2014MNRAS.441...24A,2014JCAP...05..027F} likelihoods were already interfaced with the two aforementioned MCMC codes. The SZ likelihood \citep{2014A&A...571A..20P} is present in the CosmoMC package, while we implemented it ourselves in the Monte Python code. We also used our dedicated module for computing the likelihood associated with the X-ray cluster sample \citep{2015A&A...582A..79I}. While massive neutrinos are already implemented in both the CLASS and CAMB codes, we implemented there the other $\Lambda$CDM extension we considered, namely, the $\gamma$ model. Additionally, we modified our X-ray and SZ clusters likelihood modules accordingly to support the massive neutrinos case and $\gamma$ parametrisation, all according to the recipes described earlier.

2D confidence regions and 1D constraints on parameters were then derived (as is usual in a Bayesian formalism) from the posterior distribution resulting from the MCMC chains. All parameters limits mentioned in the text are to be read as 68\% confidence limits.

\section{Results}\label{sec:3-Res}

\begin{figure}[t]
    \centering
    \includegraphics[width=\columnwidth]{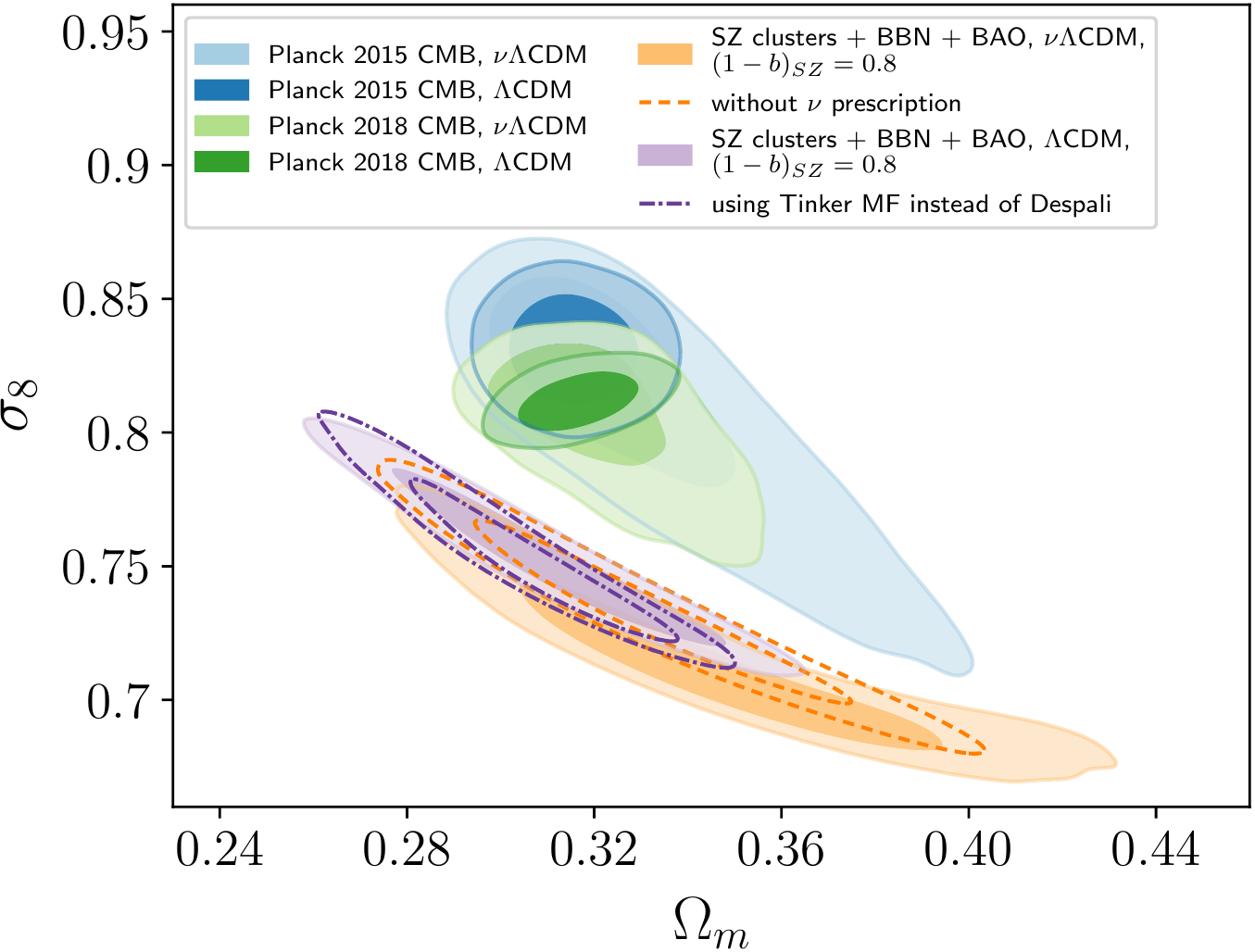}
    \caption{ Confidence contours (68 and 95\%) in the $\Omega_m-\sigma_8$ plane for various datasets: Planck 2015 and 2018 CMB data, and Planck 2015 SZ cluster counts (using the D16 mass function). Contours are shown for the minimal neutrino case and three neutrinos with free masses (using the CDM prescription for cluster counts). For SZ clusters, the calibration parameter $(1-b)$ is fixed to its fiducial value of 0.8, with priors added to $\Omega_b$ from BBN constraints, and to $H_0$ from BAO data. Two additional dashed and dash-dotted contours are shown to illustrate  the effects of changing the mass function used (from \citealt{2016MNRAS.456.2486D} to \citealt{2008ApJ...688..709T}) and removing the CDM prescription, respectively.}
    \label{fig:om_vs_s8} 
\end{figure}

The long-standing, so-called discrepancy between SZ cluster counts and the Planck CMB measurements in the $\Lambda$CDM model is illustrated in Fig.~\ref{fig:om_vs_s8} in the $\Omega_m - \sigma_8$ plane, with contours from the two probes: in blue for the 2015 Planck CMB data, and purple for SZ cluster counts (the calibration parameter $(1-b)$ being fixed to the Planck Collaboration fiducial value of $0.8$). One can see that the separation between the two contours is significantly large and remains essentially identical when considering the new Planck 2018 data release (green contours). Allowing massive neutrinos with varying mass has the effect of pushing both the CMB contours (light blue and light green contours respectively for the 2015 and 2018 Planck CMB data) and the SZ cluster contours (orange) to higher values of $\Omega_m$ and lower values of $\sigma_8$, elongating the two in the direction of the $\Omega_m - \sigma_8$ degeneracy, with the CDM prescription shifting  the contours slightly to higher $\Omega_m $ and lower $\sigma_8$.

The SZ clusters-CMB tension is therefore not relieved by the possible presence of massive neutrinos, leading to essentially identical conclusions in the case where only a sample of local X-ray clusters are used \citep{2018A&A...620A..78S}. It should be noted that in Fig.~\ref{fig:om_vs_s8} we added constraints from Big Bang nucleosynthesis (BBN) and BAO data to restrict $\Omega_b$ and $H_0$ to realistic values for cluster contours. We checked that considering wider priors results in more elongated contours in the direction of the $\Omega_m - \sigma_8$ degeneracy, but without relieving the tension.

\subsection{The \texorpdfstring{$\mathit{\Lambda}$}{L}CDM case}\label{ssec:3.1-Lcdm}

\begin{figure}[t]
    \centering
    \includegraphics[width=\columnwidth]{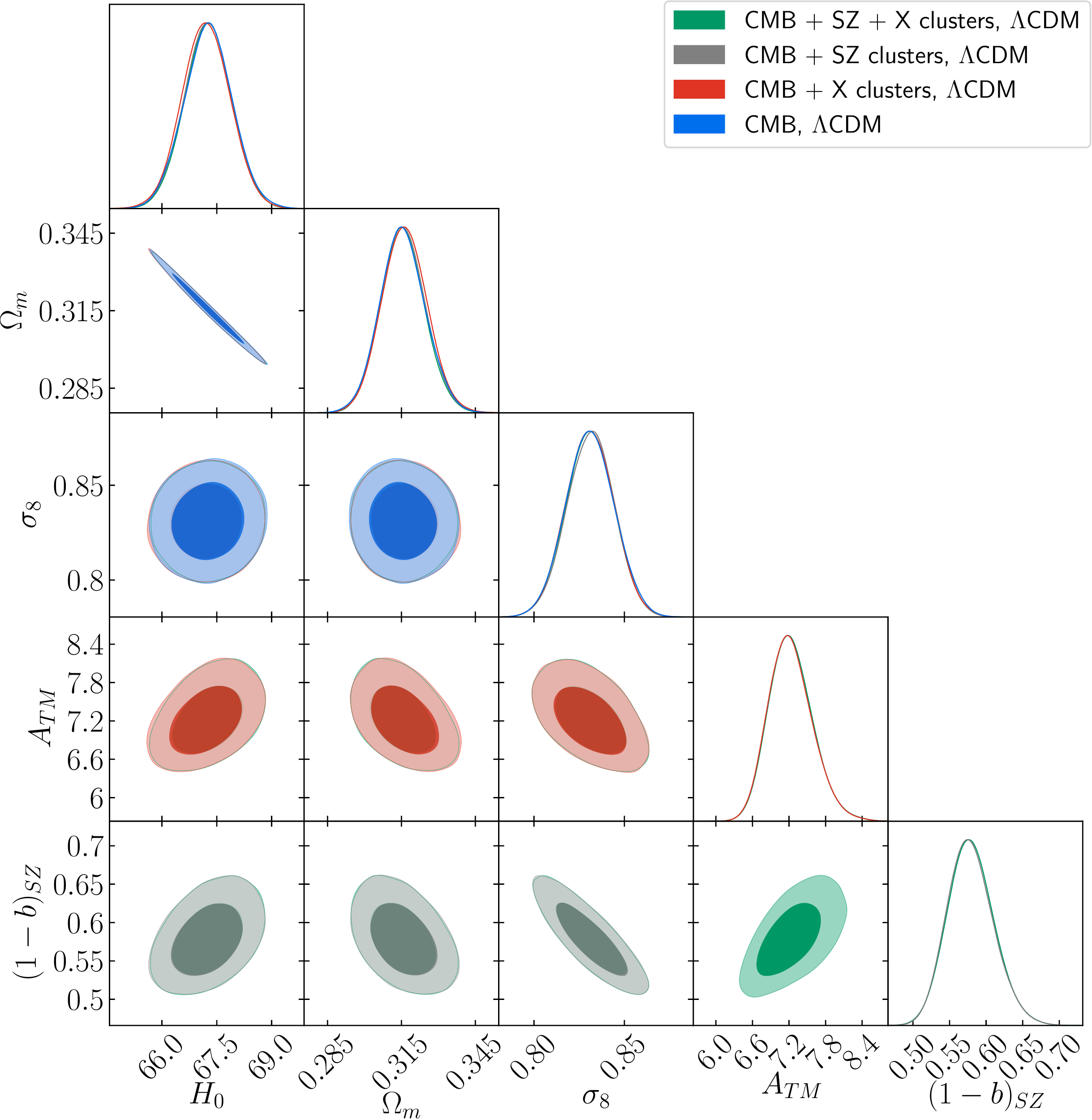}
    \caption{Confidence contours (68 and 95\%) for cosmological parameters $\Omega_m$ and $\sigma_8$, as well as mass calibration parameters $(1-b)$ and $A_{T-M}$ when combining CMB data with X-ray and/or SZ clusters data for D16 mass function.}
    \label{fig:cosmo_tri}
\end{figure}

In order to examine this tension in a quantitative way, we combine CMB data and the SZ cluster sample, adding or not adding the constraints obtained with the X-ray cluster sample, and freeing the SZ and X-ray calibration factors, as well as the slope of the mass-SZ relation and its dispersion. In the standard $\Lambda$CDM model (as defined by the Planck Collaboration, which includes one massive neutrino with a minimal mass of 0.06 eV) one can see from Fig.~\ref{fig:cosmo_tri} that, in the cases of CMB data with SZ and/or X-ray clusters, the constraints on all cosmological parameters are essentially identical. We obtain $(1-b) = 0.58\pm 0.03$ and note a clear correlation between the $(1-b)$ calibration factor for SZ clusters and the $A_{T-M}$ calibration for X-ray clusters (at virial mass), indicating the consistency of the constraints obtained from the two types of cluster samples.

\subsection{Are massive neutrinos a solution?}\label{ssec:3.2-ResNu}

\begin{figure}[t]
    \centering
    \includegraphics[width=\columnwidth]{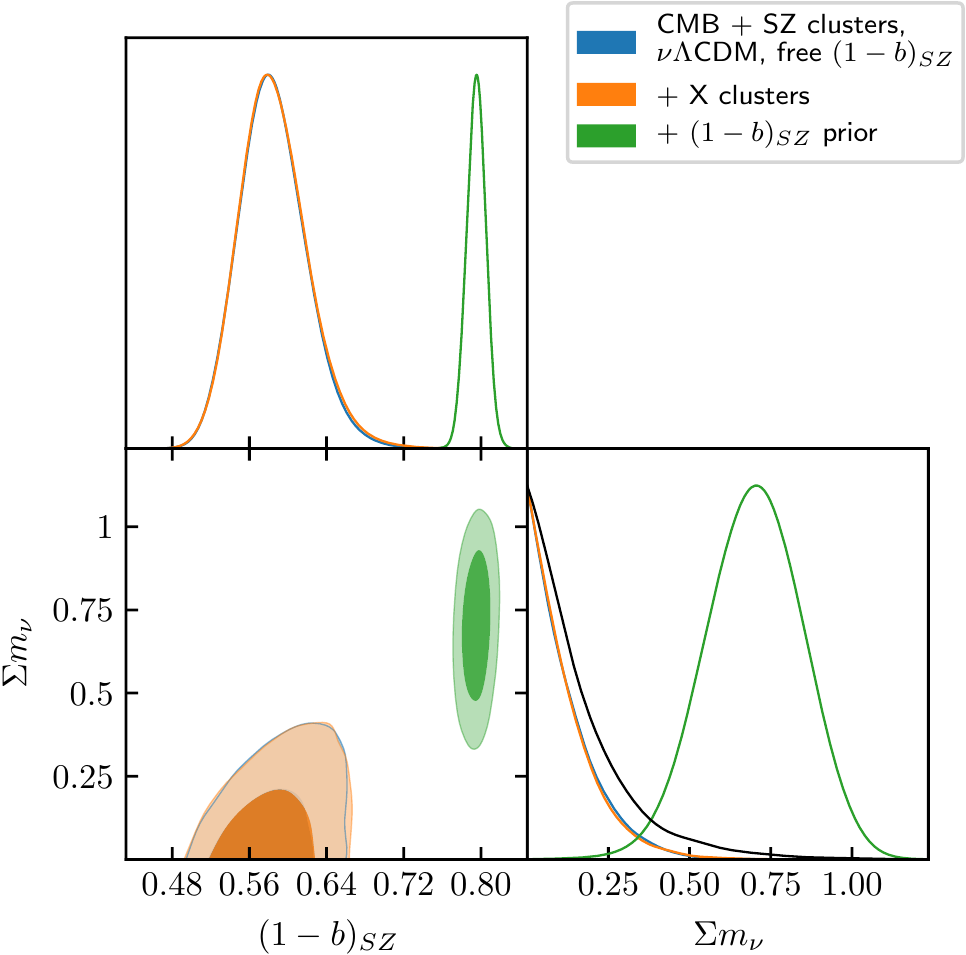}
    \caption{Confidence contours (68 and 95\%) and posterior distributions for neutrino masses $\sum m_\nu$ and mass calibration parameter $(1-b)$ when combining CMB and SZ cluster data, for D16 mass function (blue). The effect of adding X-ray clusters data is shown in orange, and the effect of a strong prior on $(1-b)$ centred around the Planck calibration value of 0.8 in green. For comparison, the 1D posterior on neutrino masses from Planck CMB data alone is also shown in black.}
    \label{fig:tri_nuLCDM_sumnu}
\end{figure}

The possible role of massive neutrinos in alleviating the cluster-CMB tension is discussed in past literature, with somewhat different conclusions, \citep{2013arXiv1302.5086R,2014PhRvD..90h3503D, 2015MNRAS.446.2205M,2015MNRAS.447.1761R,2016PDU....12...56B, 2018A&A...614A..13S} often without clear indication as to whether conclusions depend on a calibration choice. In Fig.~\ref{fig:tri_nuLCDM_sumnu}, we present the constraints on the SZ cluster calibration parameter $(1-b)$ from the combination of CMB data sets with the SZ cluster sample in the presence of massive neutrinos (blue contours). We obtain a constraint on the calibration $(1-b) = 0.585^{+0.031}_{-0.037}$ essentially identical to the minimal neutrino case. Furthermore, no strong correlation appears in the calibration-neutrino masses plane, similarly to what is obtained from the X-ray sample alone \citep{2018A&A...620A..78S}. We also notice that the constraint on neutrino masses is quite noticeably improved by combining clusters with the CMB (with 68\% limits being reduced by about $\sim 28\%$) compared to the CMB-only constraint (black curve). The standard Planck mass calibration ($(1-b) \sim 0.8$) is formally rejected at more than 5 $\sigma$. However, for illustration purposes, we show the consequence of putting a strong (Gaussian) prior on $(1-b)$ consistent with the standard Planck calibration, namely $(1-b) = 0.8 \pm 0.01$ (green contours). In this case, the favoured neutrino masses are found to be non-zero: $\sum m_\nu = 0.70\pm 0.15 $ eV, but the corresponding best-fit model has a $\Delta \chi^2 \approx 20$ compared to the case without a prior on $(1-b)$. This result is consistent with previous investigations obtaining non-zero neutrino masses using such a prior. It is therefore important to realise that Planck CMB and Planck SZ cluster counts are perfectly consistent with one another, but as mentioned before, this combination of data requires an SZ mass calibration of $(1-b) \sim 0.6$, significantly lower than the standard calibration $(1-b) \sim 0.8$.

\subsection{Introducing modified gravity }\label{ssec:3.3-ResGamma}

If the standard calibration $(1-b) \sim 0.8$ is consolidated by observations, the conclusion of the previous sections leads us to examine alternative theories to $\Lambda$CDM. One option is to consider a variable dark energy equation of state $w$ instead of a simple cosmological constant. However, \citet{2018A&A...614A..13S} found that a constant $w$ does not lead to a significant change in the required calibration. Any proposed alternative explanation should lead somehow to a lower amplitude of matter fluctuations at low redshift without altering the remarkable agreement with the CMB seen in the standard cosmological picture. Such a lower $\sigma_8$ would naturally come from a late-time deviation of the growth of matter fluctuations with respect to $\Lambda$CDM, a possibility that could naturally arise from a modification of the theory of general relativity for gravity. In the following, we will consider a phenomenological approach through the introduction of the growth index $\gamma$ following Eq.~(\ref{eq:omgam}). We follow the same approach as in previous sections and perform an MCMC analysis to explore the space of the standard cosmological parameters, adding the calibration $(1-b)$ and the index $\gamma$ as additional free parameters under the constraints of our data sets. As previously found in \citet{2018A&A...620A..78S}, we expect increasing $\gamma$ to yield a lower $\sigma_8,$ and thereby a higher calibration $(1-b)$ - corresponding to clusters being less massive. Similarly to the previous section, we use the combination of CMB and SZ cluster data, with or without adding our X-ray cluster sample to constrain the parameters of the model.

\begin{figure}[t]
    \centering
    \includegraphics[width=\columnwidth]{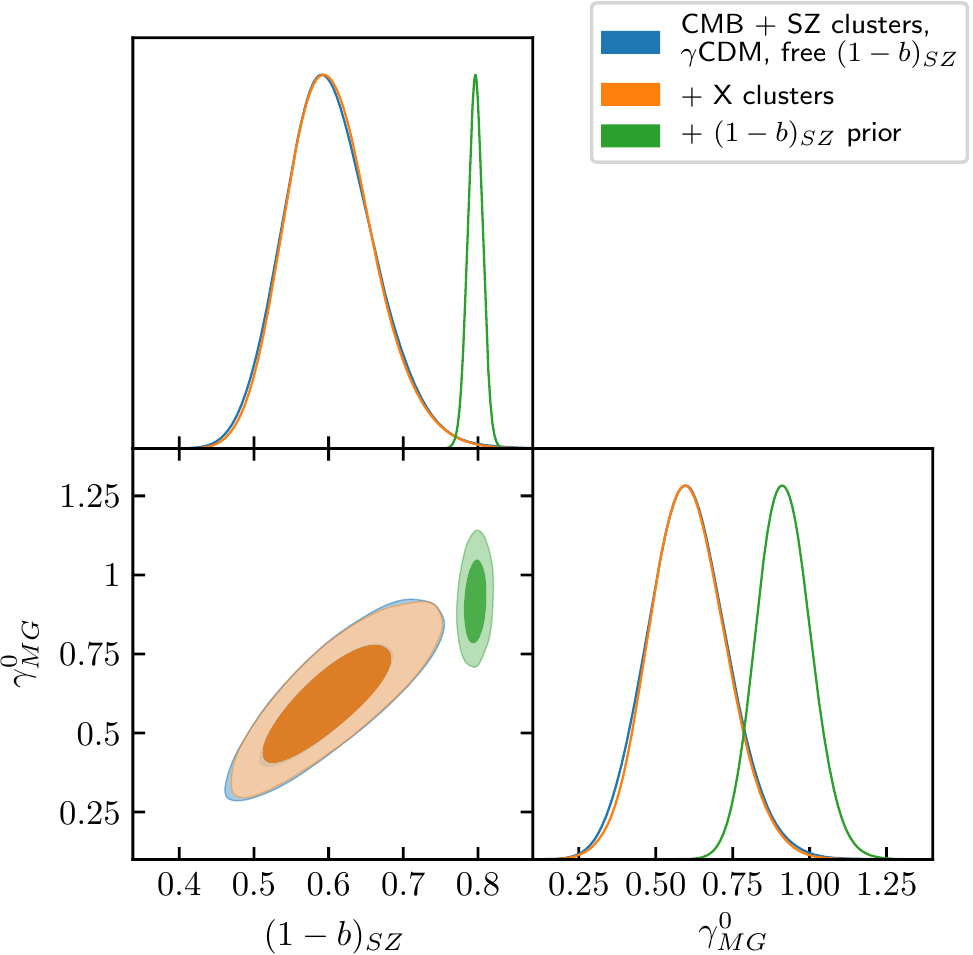}
    \caption{Confidence contours (68 and 95\%) and posterior distributions for growth index $\gamma$ and mass calibration parameter $(1-b)$ when combining CMB and SZ cluster data for D16 mass function (blue). The effect of adding X-ray clusters data is shown in orange, and the effect of a strong prior on $(1-b)$ centred around the Planck calibration value of 0.8 in green.}
    \label{fig:tri_gamCDM_gam}
\end{figure}

Our results are summarised in Fig.~\ref{fig:tri_gamCDM_gam}; the SZ calibration parameter $(1-b)$ shows a clear correlation with $\gamma$, as anticipated. However, a significant difference appears compared to the case where only the X-ray sample is used: as the SZ sample spans a wide range of (comparatively) higher redshifts, it provides additional constraints on the growth rate. As a consequence, more extreme values of $\gamma$ are rejected by the SZ cluster data, resulting in closed contours in the $\gamma$ versus $(1-b)$ plane (blue contours). The situation remains essentially unchanged when we add the X-ray sample data (orange contours). The constraint on the calibration reads $(1-b)= 0.602^{+0.053}_{-0.065}$, and the standard calibration $(1-b)=0.8$ is rejected at more than 99\%, while $\gamma = 0.60\pm 0.13$ is consistent with the growth rate in the standard $\Lambda$CDM model ($\sim0.545$). This illustrates the fact that even a simple modified version of the growth rate cannot easily accommodate  the calibration $(1-b) = 0.8$. Again for illustration purposes, we show the consequences of putting a strong prior on $(1-b),$ consistent with standard Planck calibration ($=0.8 \pm 0.01$, green contours). The value of $\gamma$ is then found to be constrained to $\gamma = 0.917 \pm 0.088,$ consistent with conclusions derived from the X-ray sample alone \citep{2018A&A...620A..78S}.

\subsection{Allowing massive neutrinos and modified gravity }\label{ssec:3.4-ResNuGamma}

\begin{figure}[t]
    \centering
    \includegraphics[width=\columnwidth]{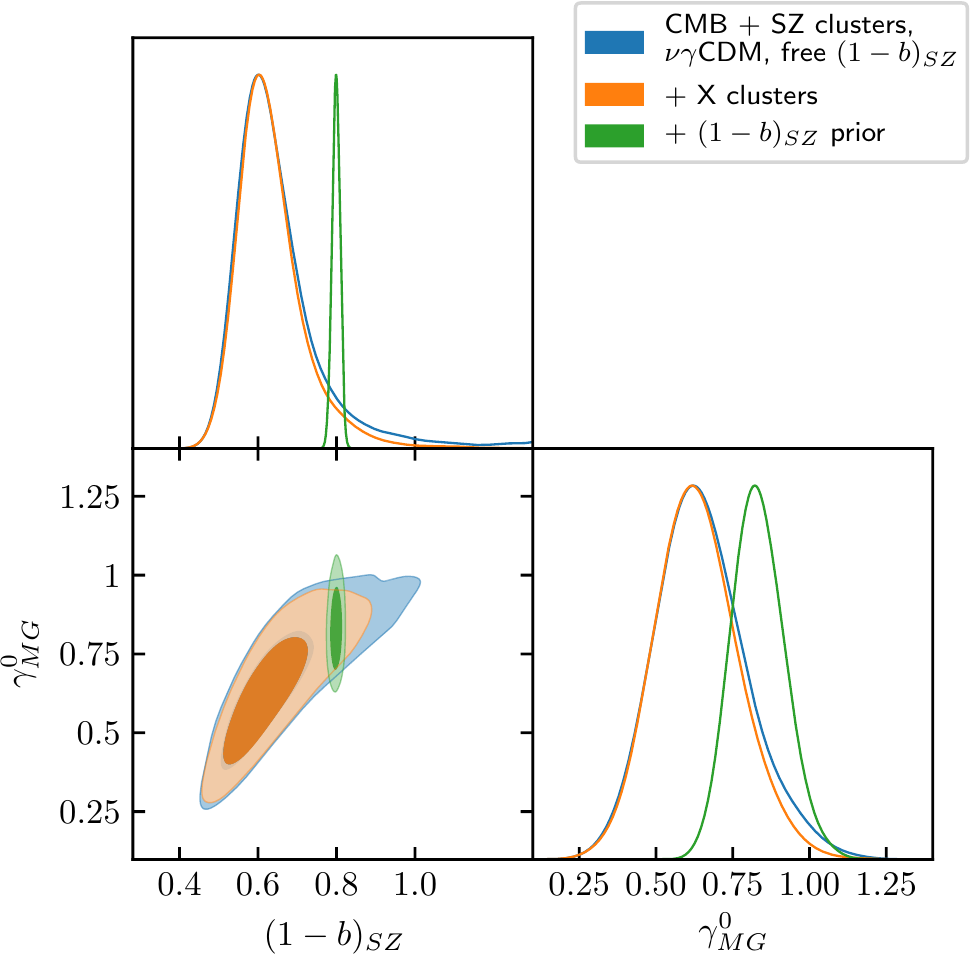}
    \caption{Confidence contours (68 and 95\%) and posterior distributions for growth index $\gamma$ and mass calibration parameter $(1-b)$ when combining CMB and SZ clusters data for D16 mass function (blue). The effect of adding X-ray clusters data is shown in orange, and the effect of a strong prior on $(1-b)$ centred around the Planck calibration value of 0.8 in green.}
    \label{fig:tri_nugamCDM_gam}
\end{figure}

\begin{figure}[t]
    \centering
    \includegraphics[width=\columnwidth]{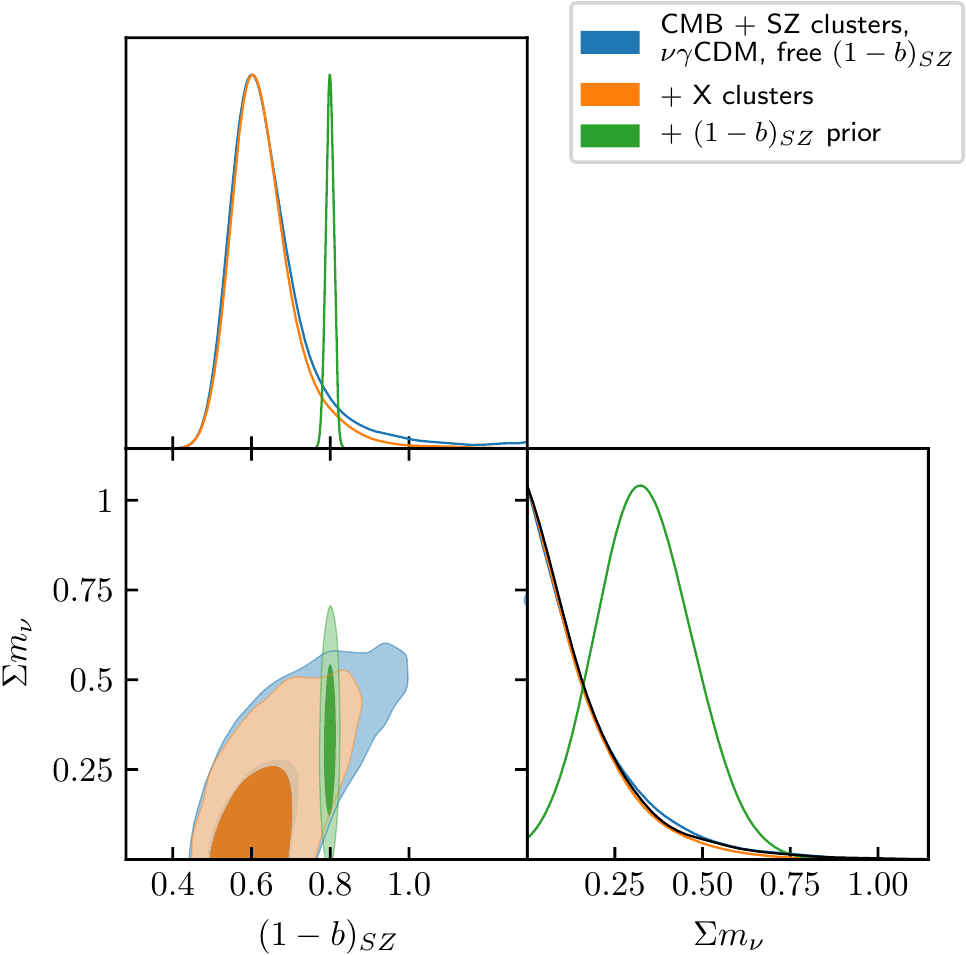}
    \caption{Confidence contours (68 and 95\%) and posterior distributions for neutrino masses $\sum m_\nu$ and mass calibration parameter $(1-b)$ when combining CMB and SZ clusters data. The legend and colours are identical to Fig.~\ref{fig:tri_nugamCDM_gam}. For comparison, the 1D posterior on neutrino masses from Planck CMB data alone is also shown in black.}
    \label{fig:tri_nugamCDM_sumnu}
\end{figure}

Here, we repeat the previous analysis, allowing both massive neutrinos and a modified gravity through the growth index $\gamma$. As can be seen from Fig.~\ref{fig:tri_nugamCDM_gam} and \ref{fig:tri_nugamCDM_sumnu}, when combining CMB and SZ clusters, a degeneracy is present between the effects of massive neutrinos and modified gravity. The preferred model remains, however, close to the standard picture: $(1-b) = 0.651^{+0.043}_{-0.12}$, $\gamma = 0.65^{+0.13}_{-0.16}$ and $\sum m_\nu \leq 0.197$, although an SZ calibration of $(1-b) \sim 0.8 $ is allowed within the two $\sigma$ posterior contours. The addition of X-ray clusters reduces the contours, although in a marginal way. When constraining $(1-b) \sim 0.8$ with a strong prior, the preferred parameters are still quite far away from the standard model: $\gamma = 0.834^{+0.080}_{-0.094}$ and $\sum m_\nu = 0.34^{+0.13}_{-0.15}$ eV, but we note that a model with $\gamma \sim $0.65 and $\sum m_\nu \sim 0.1$ is within the two $\sigma$ contours.

\subsection{Freeing the amplitude of the late-time matter power spectrum}\label{ssec:3.5-ResSig8}

As a final illustration of what our current data sets actually tell us, we performed the same analysis once again using a combination of CMB and cluster data in the $\Lambda$CDM picture, but allowed the present-day amplitude of matter fluctuations $\sigma_8$ to be an additional free parameter, no longer derived from an extrapolation at late times of the primordial amplitude $A_s$. More precisely, we assume that the growth rate between redshift zero and $\sim 1$ (the redshift range containing both the SZ and X-ray cluster samples) follows the standard $\Lambda$CDM growth rate, but modified at earlier times through a process about which we do not assume anything. In practice, we rescale the $\Lambda$CDM matter power spectrum for $z \in [0, 1]$ by a constant factor, so that we obtain the desired input $\sigma_8$ while keeping the same value of the primordial $A_s$. Such an approach allows for (comparatively) more freedom than the $\gamma$ parametrisation, as the latter only produces a very specific redshift evolution for the growth rate.

\begin{figure}[t]
    \centering
    \includegraphics[width=\columnwidth]{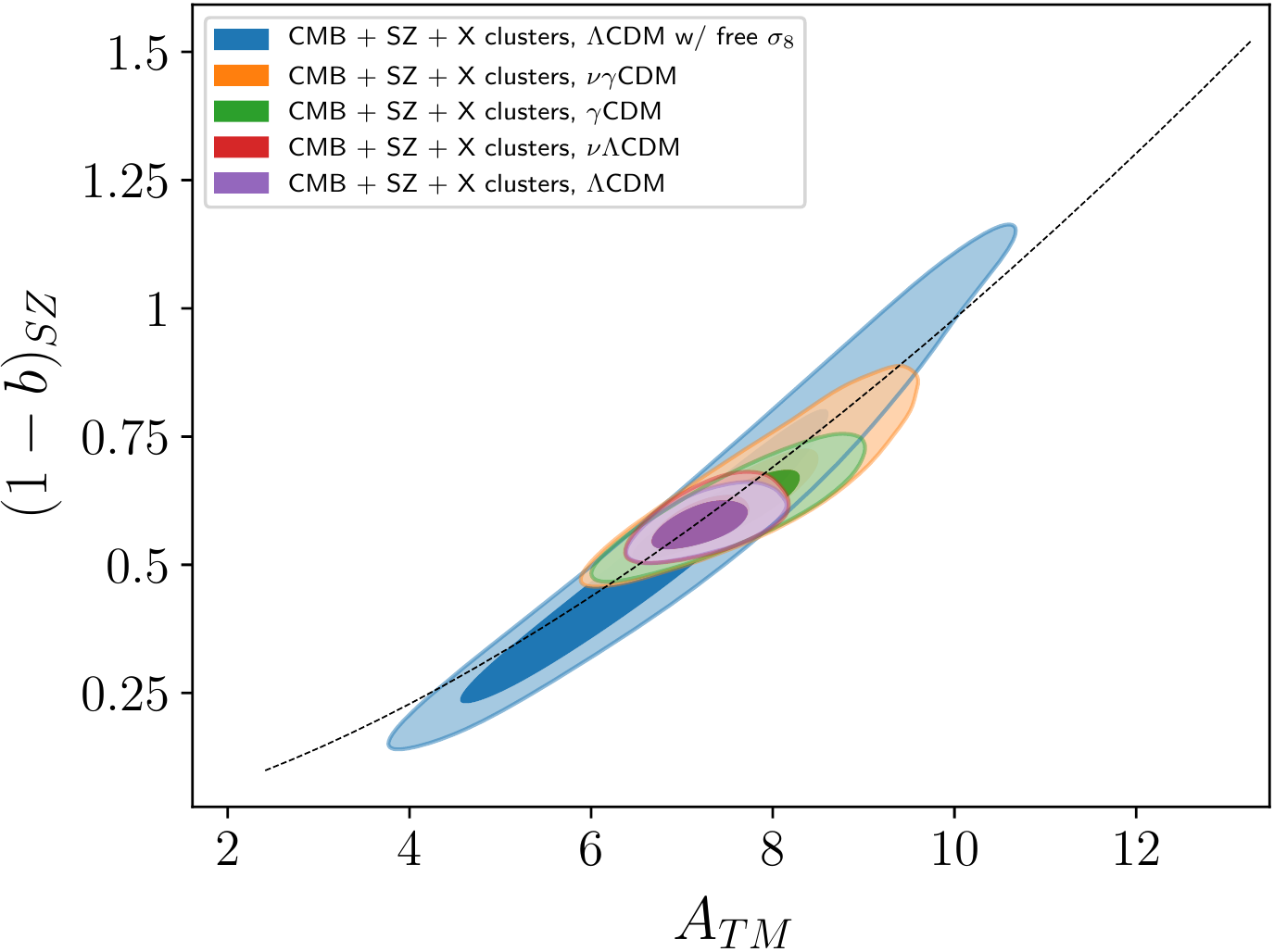}
    \caption{Confidence contours (68 and 95\%) in $(1-b)$ vs. $A_{T-M}$ plane, for combination of CMB, SZ and X-ray cluster data in various cosmological models, with or without massive neutrinos, with or without free growth index $\gamma$, and free $\sigma_8$. The black dashed line corresponds to the $(1-b)$ - $A_{T-M}$ relation leading to the same derived (SZ and X-ray) masses for a selected sample of clusters (see text for details).}
    \label{fig:atm_vs_bSZ}
\end{figure}

\begin{figure}[t]
    \centering
    \includegraphics[width=0.9\columnwidth]{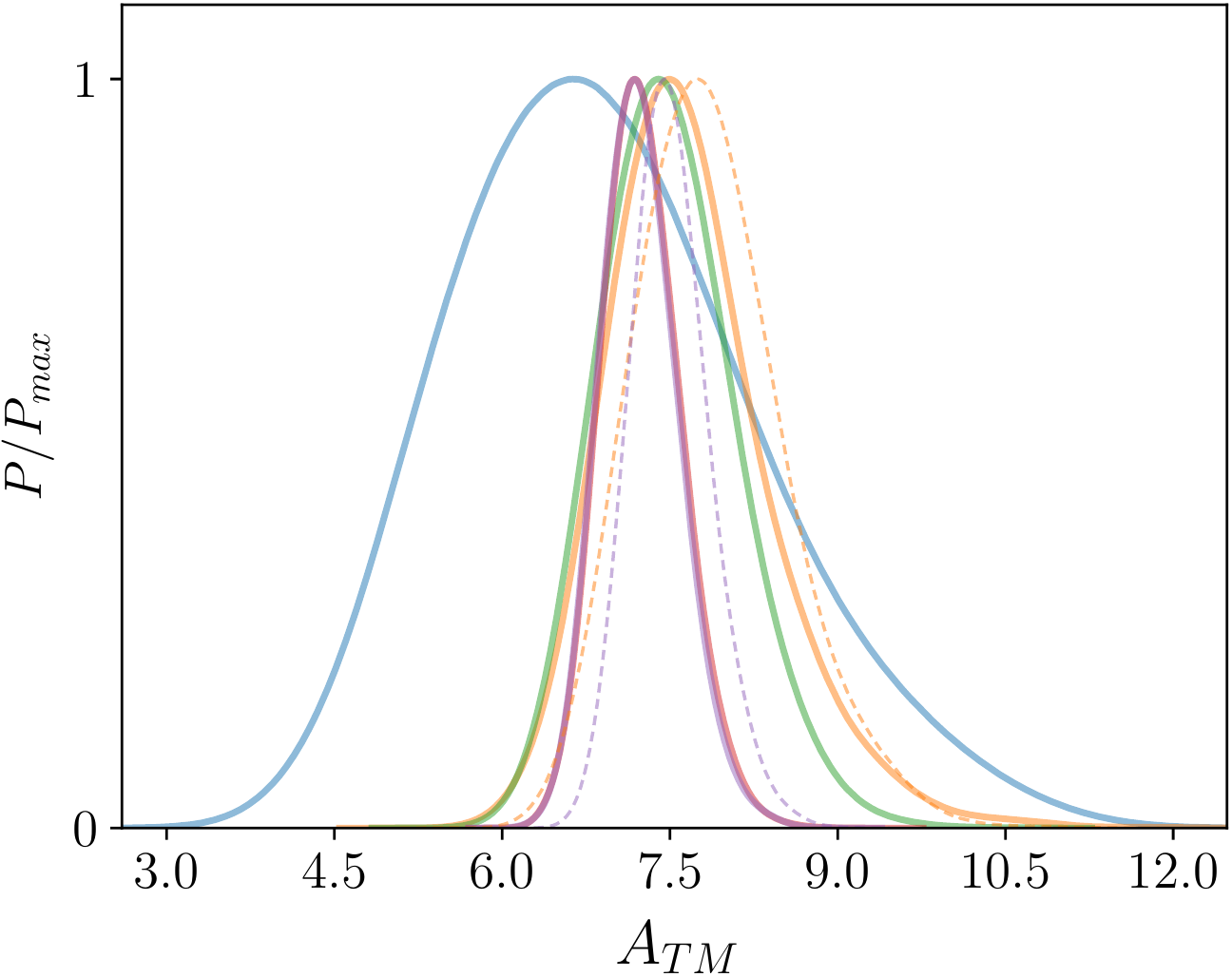} \\
    \includegraphics[width=0.9\columnwidth]{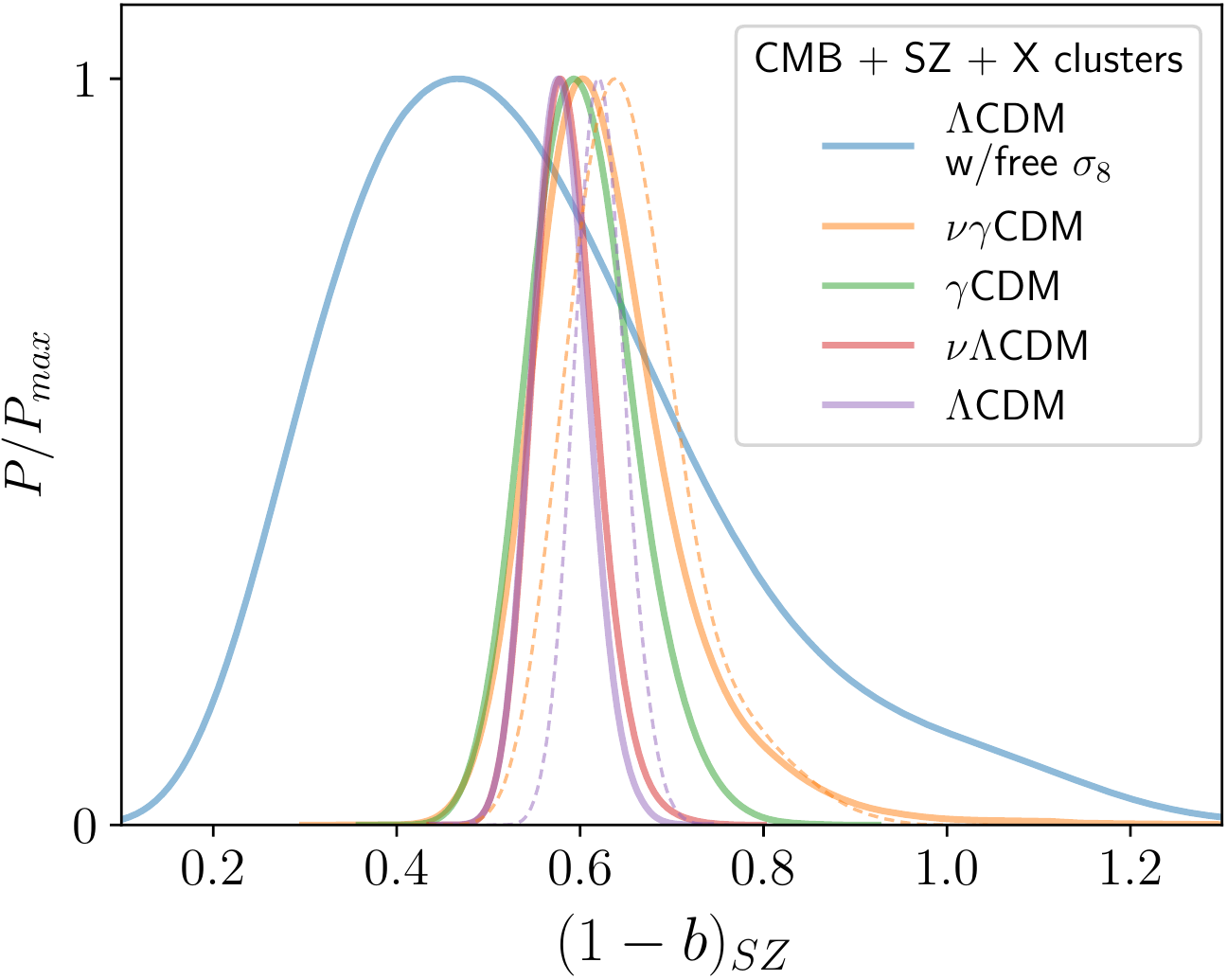}
    \caption{Marginalised posterior distributions for $A_{T-M}$ and $(1-b)$ calibration parameters, for combination of CMB, SZ and X-ray cluster data in various cosmological models: with or without massive neutrinos, with or without free growth index $\gamma$, and free $\sigma_8$. For the $\Lambda$CDM and $\nu\gamma$CDM models, the dashed lines show the same posterior obtained when using the Planck 2018 CMB data instead of the 2015 one.}
    \label{fig:atm_bSZ_like}
\end{figure}

In Fig.~\ref{fig:atm_vs_bSZ}, we show confidence contours in calibration parameters plane, for example, $(1-b)$ versus $A_{T-M}$, constrained by the combination of CMB data and our two cluster samples: X-ray and SZ. The various contours shown refer to the cases we examined in the previous sections, with the addition of the new model described in this section ($\Lambda$CDM $+$ free $\sigma_8$ at present epoch). We observe that all contours lay along the correspondence line\footnote{We derived this line using a subset of the Planck SZ clusters sample, for which both temperature and SZ measurements are available.} (dashed black) between $(1-b)$ and $A_{T-M,}$ and that in all cases the preferred SZ mass calibration $(1-b)$ remains around 0.6 (see 1D posteriors in Fig.~\ref{fig:atm_bSZ_like}) even in the case where $\sigma_8$ is treated as a free parameter. In the latter case, a value of $(1-b) \sim 0.8$ is still acceptable (meaning within reasonable confidence limits), but would thus require a significant modification of the growth rate at earlier times in order to reach the required current value of $\sigma_8$.

\section{Discussion and conclusions}\label{sec:4-CCL}

In the present work, we have examined the constraints on the calibration constants in the scaling relations of the Planck SZ and a local X-ray cluster sample in the context of various cosmological models and assumptions. In our approach, we treat these calibrations (namely $(1-b)$ for SZ clusters, and $A_{T-M}$ for X-ray clusters) as free parameters without any priors. Neither massive neutrinos nor a modified gravity model (parametrised by the $\gamma$ growth index) allow us to reconcile the Planck data (CMB and SZ clusters) with the standard SZ mass calibration value of $(1-b)\sim 0.8$. Consequently, this implies that there is no simple solution to the so-called clusters-CMB tension, which may be more accurately described as a tension between Planck data and the empirical calibration of the mass-SZ observable (which yields the $(1-b)\sim 0.8$ value). Despite this observation, we explore the possibility of more extended models allowing massive neutrinos, and a modification of gravity, as well as the approach of freeing the amplitude of matter fluctuations at low redshift (as measured by $\sigma_8$). Nevertheless, the SZ cluster counts combined with CMB data still prefer a low calibration ($(1-b)\sim 0.6$) in all considered models. We therefore find it striking that no cosmological model appears to prefer the (comparatively) high calibration of $(1-b)\sim 0.8$.

On a final note, this work was finished within a week of the Planck 2018 CMB likelihood being published, hence the use of the older 2015 data instead. We did check, however, that for the $\Lambda$CDM case notably, using this new CMB data yields a slight difference on the preferred values of $(1-b) = 0.622\pm 0.028,$ which still leaves the level of tension above $6 \sigma$ (see Fig.~\ref{fig:atm_bSZ_like}). Therefore, we do not expect significant changes to the other results of our study and to our conclusions.

\begin{acknowledgements}

SI was supported by the European Structural and Investment Fund and the Czech Ministry of Education, Youth and Sports (Project CoGraDS - CZ.02.1.01/0.0/0.0/15\_003/0000437). ZS was supported by a grant of excellence from the Agence des Universités Francophones (AFU). This work has been carried out thanks to the support of the OCEVU Labex (ANR-11-LABX-0060) and the A*MIDEX project (ANR-11-IDEX-0001-02) funded by the ``Investissements d'Avenir'' French government program managed by the ANR. It was also conducted using resources from the IN2P3 Lyon computing center. We thank the referee for their relevant comments and recommendations.

\end{acknowledgements}

\bibliographystyle{aa}
\bibliography{references}

\begin{thebibliography}{29}
\expandafter\ifx\csname natexlab\endcsname\relax\def\natexlab#1{#1}\fi

\bibitem[{{Anderson} {et~al.}(2014){Anderson}, {Aubourg}, {Bailey}, {Beutler},
  {Bhardwaj}, {Blanton}, {Bolton}, {Brinkmann}, {Brownstein}, {Burden},
  {Chuang}, {Cuesta}, {Dawson}, {Eisenstein}, {Escoffier}, {Gunn}, {Guo}, {Ho},
  {Honscheid}, {Howlett}, {Kirkby}, {Lupton}, {Manera}, {Maraston}, {McBride},
  {Mena}, {Montesano}, {Nichol}, {Nuza}, {Olmstead}, {Padmanabhan},
  {Palanque-Delabrouille}, {Parejko}, {Percival}, {Petitjean}, {Prada},
  {Price-Whelan}, {Reid}, {Roe}, {Ross}, {Ross}, {Sabiu}, {Saito}, {Samushia},
  {S{\'a}nchez}, {Schlegel}, {Schneider}, {Scoccola}, {Seo}, {Skibba},
  {Strauss}, {Swanson}, {Thomas}, {Tinker}, {Tojeiro}, {Maga{\~n}a}, {Verde},
  {Wake}, {Weaver}, {Weinberg}, {White}, {Xu}, {Y{\`e}che}, {Zehavi}, \&
  {Zhao}}]{2014MNRAS.441...24A}
{Anderson}, L., {Aubourg}, {\'E}., {Bailey}, S., {et~al.} 2014, \mnras, 441, 24

\bibitem[{{Arnaud} {et~al.}(2010){Arnaud}, {Pratt}, {Piffaretti},
  {B{\"o}hringer}, {Croston}, \& {Pointecouteau}}]{2010A&A...517A..92A}
{Arnaud}, M., {Pratt}, G.~W., {Piffaretti}, R., {et~al.} 2010, \aap, 517, A92

\bibitem[{Audren {et~al.}(2013)Audren, Lesgourgues, Benabed, \&
  Prunet}]{Audren:2012wb}
Audren, B., Lesgourgues, J., Benabed, K., \& Prunet, S. 2013, JCAP, 1302, 001

\bibitem[{{Bull} {et~al.}(2016){Bull}, {Akrami}, {Adamek}, {Baker}, {Bellini},
  {Beltr{\'a}n Jim{\'e}nez}, {Bentivegna}, {Camera}, {Clesse}, {Davis}, {Di
  Dio}, {Enander}, {Heavens}, {Heisenberg}, {Hu}, {Llinares}, {Maartens},
  {M{\"o}rtsell}, {Nadathur}, {Noller}, {Pasechnik}, {Pawlowski}, {Pereira},
  {Quartin}, {Ricciardone}, {Riemer-S{\o}rensen}, {Rinaldi}, {Sakstein},
  {Saltas}, {Salzano}, {Sawicki}, {Solomon}, {Spolyar}, {Starkman}, {Steer},
  {Tereno}, {Verde}, {Villaescusa-Navarro}, {von Strauss}, \&
  {Winther}}]{2016PDU....12...56B}
{Bull}, P., {Akrami}, Y., {Adamek}, J., {et~al.} 2016, Physics of the Dark
  Universe, 12, 56

\bibitem[{{Castorina} {et~al.}(2014){Castorina}, {Sefusatti}, {Sheth},
  {Villaescusa-Navarro}, \& {Viel}}]{2014JCAP...02..049C}
{Castorina}, E., {Sefusatti}, E., {Sheth}, R.~K., {Villaescusa-Navarro}, F., \&
  {Viel}, M. 2014, \jcap, 2, 049

\bibitem[{{Clifton} {et~al.}(2012){Clifton}, {Ferreira}, {Padilla}, \&
  {Skordis}}]{2012PhR...513....1C}
{Clifton}, T., {Ferreira}, P.~G., {Padilla}, A., \& {Skordis}, C. 2012,
  \physrep, 513, 1

\bibitem[{{Costanzi} {et~al.}(2013){Costanzi}, {Villaescusa-Navarro}, {Viel},
  {Xia}, {Borgani}, {Castorina}, \& {Sefusatti}}]{2013JCAP...12..012C}
{Costanzi}, M., {Villaescusa-Navarro}, F., {Viel}, M., {et~al.} 2013, \jcap,
  12, 012

\bibitem[{{Despali} {et~al.}(2016){Despali}, {Giocoli}, {Angulo}, {Tormen},
  {Sheth}, {Baso}, \& {Moscardini}}]{2016MNRAS.456.2486D}
{Despali}, G., {Giocoli}, C., {Angulo}, R.~E., {et~al.} 2016, \mnras, 456, 2486

\bibitem[{{Dvorkin} {et~al.}(2014){Dvorkin}, {Wyman}, {Rudd}, \&
  {Hu}}]{2014PhRvD..90h3503D}
{Dvorkin}, C., {Wyman}, M., {Rudd}, D.~H., \& {Hu}, W. 2014, \prd, 90, 083503

\bibitem[{{Font-Ribera} {et~al.}(2014){Font-Ribera}, {Kirkby}, {Busca},
  {Miralda-Escud{\'e}}, {Ross}, {Slosar}, {Rich}, {Aubourg}, {Bailey},
  {Bhardwaj}, {Bautista}, {Beutler}, {Bizyaev}, {Blomqvist}, {Brewington},
  {Brinkmann}, {Brownstein}, {Carithers}, {Dawson}, {Delubac}, {Ebelke},
  {Eisenstein}, {Ge}, {Kinemuchi}, {Lee}, {Malanushenko}, {Malanushenko},
  {Marchante}, {Margala}, {Muna}, {Myers}, {Noterdaeme}, {Oravetz},
  {Palanque-Delabrouille}, {P{\^a}ris}, {Petitjean}, {Pieri}, {Rossi},
  {Schneider}, {Simmons}, {Viel}, {Yeche}, \& {York}}]{2014JCAP...05..027F}
{Font-Ribera}, A., {Kirkby}, D., {Busca}, N., {et~al.} 2014, \jcap, 5, 027

\bibitem[{{Ili{\'c}} {et~al.}(2015){Ili{\'c}}, {Blanchard}, \&
  {Douspis}}]{2015A&A...582A..79I}
{Ili{\'c}}, S., {Blanchard}, A., \& {Douspis}, M. 2015, \aap, 582, A79

\bibitem[{{Ishak}(2019)}]{2019LRR....22....1I}
{Ishak}, M. 2019, Living Reviews in Relativity, 22, 1

\bibitem[{{Lesgourgues} \& {Pastor}(2012)}]{2012arXiv1212.6154L}
{Lesgourgues}, J. \& {Pastor}, S. 2012, ArXiv e-prints

\bibitem[{{Lewis}(2013)}]{2013PhRvD..87j3529L}
{Lewis}, A. 2013, \prd, 87, 103529

\bibitem[{{Lewis} \& {Bridle}(2002)}]{2002PhRvD..66j3511L}
{Lewis}, A. \& {Bridle}, S. 2002, \prd, 66, 103511

\bibitem[{{L'Huillier} {et~al.}(2018){L'Huillier}, {Shafieloo}, \&
  {Kim}}]{2018MNRAS.476.3263L}
{L'Huillier}, B., {Shafieloo}, A., \& {Kim}, H. 2018, \mnras, 476, 3263

\bibitem[{{Linder}(2005)}]{2005PhRvD..72d3529L}
{Linder}, E.~V. 2005, \prd, 72, 043529

\bibitem[{{Mantz} {et~al.}(2015){Mantz}, {von der Linden}, {Allen},
  {Applegate}, {Kelly}, {Morris}, {Rapetti}, {Schmidt}, {Adhikari}, {Allen},
  {Burchat}, {Burke}, {Cataneo}, {Donovan}, {Ebeling}, {Shandera}, \&
  {Wright}}]{2015MNRAS.446.2205M}
{Mantz}, A.~B., {von der Linden}, A., {Allen}, S.~W., {et~al.} 2015, \mnras,
  446, 2205

\bibitem[{{Peebles}(1980)}]{1980lssu.book.....P}
{Peebles}, P.~J.~E. 1980, {The large-scale structure of the universe}
  (Princeton University Press)

\bibitem[{{Planck Collaboration} {et~al.}(2016{\natexlab{a}}){Planck
  Collaboration}, {Adam}, {Ade}, {Aghanim}, {Akrami}, {Alves}, {Arg{\"u}eso},
  {Arnaud}, {Arroja}, {Ashdown}, \& et~al.}]{2016A&A...594A...1P}
{Planck Collaboration}, {Adam}, R., {Ade}, P.~A.~R., {et~al.}
  2016{\natexlab{a}}, \aap, 594, A1

\bibitem[{{Planck Collaboration} {et~al.}(2014){Planck Collaboration}, {Ade},
  {Aghanim}, {Armitage-Caplan}, {Arnaud}, {Ashdown}, {Atrio-Barandela},
  {Aumont}, {Baccigalupi}, {Banday}, \& et~al.}]{2014A&A...571A..20P}
{Planck Collaboration}, {Ade}, P.~A.~R., {Aghanim}, N., {et~al.} 2014, \aap,
  571, A20

\bibitem[{{Planck Collaboration} {et~al.}(2016{\natexlab{b}}){Planck
  Collaboration}, {Ade}, {Aghanim}, {Arnaud}, {Ashdown}, {Aumont},
  {Baccigalupi}, {Banday}, {Barreiro}, {Bartlett}, \&
  et~al.}]{2016A&A...594A..24P}
{Planck Collaboration}, {Ade}, P.~A.~R., {Aghanim}, N., {et~al.}
  2016{\natexlab{b}}, \aap, 594, A24

\bibitem[{{Planck Collaboration} {et~al.}(2016{\natexlab{c}}){Planck
  Collaboration}, {Aghanim}, {Arnaud}, {Ashdown}, {Aumont}, {Baccigalupi},
  {Banday}, {Barreiro}, {Bartlett}, {Bartolo}, \& et~al.}]{2016A&A...594A..11P}
{Planck Collaboration}, {Aghanim}, N., {Arnaud}, M., {et~al.}
  2016{\natexlab{c}}, \aap, 594, A11

\bibitem[{{Roncarelli} {et~al.}(2015){Roncarelli}, {Carbone}, \&
  {Moscardini}}]{2015MNRAS.447.1761R}
{Roncarelli}, M., {Carbone}, C., \& {Moscardini}, L. 2015, Monthly Notices of
  the Royal Astronomical Society, 447, 1761

\bibitem[{{Rozo} {et~al.}(2013){Rozo}, {Rykoff}, {Bartlett}, \&
  {Evrard}}]{2013arXiv1302.5086R}
{Rozo}, E., {Rykoff}, E.~S., {Bartlett}, J.~G., \& {Evrard}, A.~E. 2013, arXiv
  e-prints, arXiv:1302.5086

\bibitem[{{Sakr} {et~al.}(2018){Sakr}, {Ili{\'c}}, {Blanchard}, {Bittar}, \&
  {Farah}}]{2018A&A...620A..78S}
{Sakr}, Z., {Ili{\'c}}, S., {Blanchard}, A., {Bittar}, J., \& {Farah}, W. 2018,
  \aap, 620, A78

\bibitem[{{Salvati} {et~al.}(2018){Salvati}, {Douspis}, \&
  {Aghanim}}]{2018A&A...614A..13S}
{Salvati}, L., {Douspis}, M., \& {Aghanim}, N. 2018, \aap, 614, A13

\bibitem[{{Tinker} {et~al.}(2008){Tinker}, {Kravtsov}, {Klypin}, {Abazajian},
  {Warren}, {Yepes}, {Gottl{\"o}ber}, \& {Holz}}]{2008ApJ...688..709T}
{Tinker}, J., {Kravtsov}, A.~V., {Klypin}, A., {et~al.} 2008, The Astrophysical
  Journal, 688, 709

\bibitem[{{Zarrouk} {et~al.}(2018){Zarrouk}, {Burtin}, {Gil-Mar{\'\i}n},
  {Ross}, {Tojeiro}, {P{\^a}ris}, {Dawson}, {Myers}, {Percival}, {Chuang},
  {Zhao}, {Bautista}, {Comparat}, {Gonz{\'a}lez-P{\'e}rez}, {Habib},
  {Heitmann}, {Hou}, {Laurent}, {Le Goff}, {Prada}, {Rodr{\'\i}guez-Torres},
  {Rossi}, {Ruggeri}, {S{\'a}nchez}, {Schneider}, {Tinker}, {Wang},
  {Y{\`e}che}, {Baumgarten}, {Brownstein}, {de la Torre}, {du Mas des
  Bourboux}, {Kneib}, {Mariappan}, {Palanque-Delabrouille}, {Peacock},
  {Petitjean}, {Seo}, \& {Zhao}}]{2018MNRAS.477.1639Z}
{Zarrouk}, P., {Burtin}, E., {Gil-Mar{\'\i}n}, H., {et~al.} 2018, Monthly
  Notices of the Royal Astronomical Society, 477, 1639

\end{thebibliography}

\end{document}